\documentclass{emulateapj}

\usepackage[dvips]{epsfig,color}

\slugcomment{}

\shorttitle{Milgrom Models}
\shortauthors{Barnes, Kosowsky, \& Sellwood}

\newcommand{\chisq}{\ensuremath{\chi^2_r}}

\newcommand{\biwt}{\ensuremath{\chi^2_b}}

\newcommand{\ignore}[1]{\relax}
\newcommand{\ml}{\ensuremath{\Upsilon}}
\newcommand{\mli}{\ensuremath{\Upsilon_I}}
\newcommand{\mld}{\ensuremath{\Upsilon_{D,I}}}
\newcommand{\mlb}{\ensuremath{\Upsilon_{B,I}}}
\newcommand{\mldz}{\ensuremath{\Upsilon_{D,0}}}

\newcommand{\eg}{{\it e.g.}}
\newcommand{\ie}{{\it i.e.}}
\newcommand{\etal}{{\it et al.}}

\begin{document}

\title{Milgrom Relation Models for Spiral Galaxies from Two-Dimensional
Velocity Maps}
\author{Eric I. Barnes}
\affil{Department of Physics, University of Wisconsin---La Crosse,
La Crosse, WI 54601}
\author{Arthur Kosowsky}
\affil{Department of Physics and Astronomy, University of Pittsburgh,
Pittsburgh, PA 15260}
\author{J.A. Sellwood}
\affil{Department of Physics and Astronomy, Rutgers 
University, Piscataway, NJ 08854}
\email{barnes.eric@uwlax.edu}
\email{kosowsky@pitt.edu}
\email{sellwood@physics.rutgers.edu}

\begin{abstract}

Using two-dimensional velocity maps and $I$-band photometry, we have
created mass models of 40 spiral galaxies using the Milgrom relation
(the basis of modified Newtonian dynamics, or MOND) to complement
previous work.  A Bayesian technique is employed to compare several
different dark matter halo models to Milgrom and Newtonian models.
Pseudo-isothermal dark matter halos provide the best statistical fits
to the data in a majority of cases, while the Milgrom relation
generally provides good fits as well.  We also find that Milgrom
models give mass-to-light ratios that roughly correlate with galaxy
color, as predicted by stellar population models. A subsample of
galaxies in the Hydra cluster follow a tight relation between
mass-to-light and color, but one that is significantly different from
relations found in previous studies.  Ruling out the Milgrom relation
with rotational kinematics is difficult due to systematic
uncertainties in the observations as well as underlying model
assumptions.  We discuss in detail two galaxies for which the Milgrom
relation appears to fail and find that relaxing the assumption of
constant stellar mass-to-light ratio can maintain Milgrom models'
viability.

\end{abstract}


\section{Introduction}

The vast majority of spiral galaxy rotation curves are flatter than
Newtonian predictions at large distances from galactic centers
\citep{sal78,rub79,sa79,ks79,sas79,vm82,roh86}.  Galaxies that do not
show this behavior often show peculiar morphologies or perturbing
interactions.  This asymptotic flatness of rotation curves, which is
not perfect \citep[for a more in-depth discussion of this point see,
\eg,][]{pss96,cgh06}, disagrees with the Newtonian prediction in which
the circular orbital speed due to the gravity from a galaxy's luminous
material should decrease at distances which enclose substantial
fractions of a galaxy's visible mass.  This type of discrepancy is
usually viewed as evidence for a non-luminous mass component, known as
a dark matter halo.  This picture of luminous matter embedded in dark
matter halos has become the paradigm of galactic structure, and recent
work involving galaxy clusters \citep{clo06,bra06} has presented
strong evidence that this accurately reflects astrophysical reality.
Since a precise mass distribution of the visible matter is difficult
to determine directly, the dark matter distribution is also not known.
Previous studies \citep[\eg,][hereafter BSK]{bac01,bsk04} have found
that the choice of dark matter distribution significantly impacts the
derived mass of the luminous material.

A different phenomenological approach to rotation curve fitting is
provided by the fitting formula originally proposed by Milgrom
\citep{mil83a,mil83b,mil83c} which he termed MOND, for ``modified
Newtonian dynamics.'' The original goal of MOND was to use only the
luminous matter to explain two widely observed properties of spiral
galaxies; asymptotic flatness of galactic rotation curves and the
well-known Tully--Fisher relation $L\simeq v^\beta$ \citep{tf77}.  A
tight relation between a galaxy's luminosity $L$ and the asymptotic
value of the galaxy's circular speed $v$ appears to be valid for all
varieties of spirals, high and low surface brightness as well as
barred and unbarred galaxies \citep{spray95,z95,cr99,court03}.
\citet{mcg00} has convincingly argued that luminosity is actually a
stand-in for total baryonic mass; \ie, the same Tully--Fisher relation
holds for both low and high surface brightness spirals if baryonic
mass (stars and gas) is used in place of luminosity.  When this
baryonic Tully-Fisher relation is examined, $M_{\rm baryonic} \propto
v^4$ \citep{mcg05}.  These observations are built into the MOND
formalism as follows.  The total acceleration due to a gravitating
object is given by the expression
\begin{equation}\label{mondeq}
\mathbf{a}\mu(a/a_0)=\mathbf{a}_N,
\end{equation}
which relates the total acceleration $\mathbf{a}$ to the Newtonian
acceleration $\mathbf{a}_N$ through a function $\mu$ satisfying the
limiting forms $\mu \to 1$ when $a \gg a_0$ and $\mu \to a/a_0$ when
$a \ll a_0$ \citep{mil83a}. The new universal constant $a_0=1.2\times
10^{-10}$ km/s$^2$ is obtained from fits to measured rotation curves
and is the same for all galaxies. The exact form of the interpolating
function is essentially unconstrained by rotation curve data;
throughout this paper we use $\mu(x)=x/\sqrt{1+x^2}$ \citep{beg91}. It
has long been realized that Equation~(\ref{mondeq}) violates
conservation of linear momentum; a more acceptable gravitational
modification alters the Poisson equation for the gravitational
potential $\phi$ \citep{bm84},
\begin{equation}\label{poisson}
\nabla\cdot\left[\mu\left(\frac{a}{a_0})\nabla\phi\right)\right].
\end{equation}
This modification of the gravitational field explicitly conserves
energy and momentum, as it is based on a Lagrangian formulation.
However, in the limited case of spiral galaxies like those considered
here, the force law modification of Equation~(\ref{mondeq}) is a
sufficiently accurate approximation.

In the inner regions of typical spiral galaxies, the characteristic
acceleration scale $a\gg a_0$ and the gravitational force reduces to
the usual Newtonian force.  In the outer regions of spiral galaxies,
$a \ll a_0$ (the so-called MOND regime) leading to the acceleration
having a 1/$r$ dependence, asymptotically flat rotation curves, and an
enclosed mass varying like the flat rotation speed to the fourth
power.  Assuming a constant mass-to-light ratio \ml\ results in the
Tully--Fisher relation.  A number of previous studies have utilized
Equation~(\ref{mondeq}) to describe rotation curves
\citep{kent87,beg91,san96,sv98,dbm98,g04}.  Their success in
accurately modeling rotation curves with one free parameter (two if a
bulge is present) rivals dark matter halo models that require at least
two additional parameters.  The empirical accomplishments suggest that
regardless of the veracity of the hypothesis that Newtonian dynamics
is modified, Equation~\ref{mondeq} (the Milgrom relation) is a
powerful and compact method for characterizing the rotational
kinematics of spiral galaxies.

In this work, we regard the Milgrom relation as an empirical fitting
formula for rotation curve data.  In particular, we make no
assumptions about properties of dark matter halos, or modifications to
gravitation, from which the Milgrom relation might arise. But as
emphasized in \citet{mcg04}, Equation~(\ref{mondeq}) is {\it the}
effective force law for the visible matter in spiral galaxies,
regardless of whether the force comes from a distribution of dark
matter or from MOND, which seems increasingly unlikely in the face of
evidence like that presented in \citet{clo06,bra06} \citep[but see
also][]{a06}.  This paper remains agnostic about the ontological
status of the Milgrom force law.  However, we note that if galactic
dark matter is taken as a certainty, then the success of the Milgrom
relation in predicting spiral galaxy rotation curves reflects an
underlying dynamical process in galaxy formation which is not well
understood and is deserving of further scrutiny.  In this work, we are
interested in determining i) the extent to which this dynamical
relation holds and ii) specific situations in which the relation might
not hold.

This paper performs Milgrom fits to the same sample of spiral galaxies
considered in BSK, and provides a statistical comparison with the dark
matter halo fits. The halo profiles considered in BSK are the familiar
\citet[][hereafter NFW]{nfw96}
\begin{equation}\label{nfwrho}
\rho_{\rm NFW}(r) = \rho_s \frac{r_s^3}{r(r+r_s)^2},
\end{equation}
where $\rho_s$ is a characteristic density and $r_s$ is a scale
length; pseudo-isothermal (PI)
\begin{equation}\label{isorho}
\rho_{\rm PI} = \frac{\rho_0}{1+(r/r_c)^2},
\end{equation}
with central density  $\rho_0$ and core radius $r_c$; and a simple
power law
\begin{equation}\label{plrho}
\rho(r)=\rho_0 \left(\frac{r}{r_l}\right)^{\alpha},
\end{equation}
along with a Newtonian stars-only fit which is useful for comparing
with both Milgrom and dark halo fits.

Sample selection, photometric and kinematic data, and modeling
procedures are discussed in \S~\ref{data}.  Along with the results of
Milgrom fits to this data (\S~\ref{parvals}), we obtain the \ml-color
relation for these models (\S~\ref{colors}).  This sample of galaxies
contains 8 Hydra cluster galaxies, and we focus on these galaxies to
investigate the claim that Milgrom models produce \ml\ values that
closely agree with population synthesis predictions and therefore
provide likely estimates of luminous mass.  We also compare the
Milgrom model fits to the dark halo fits from BSK, utilizing a
Bayesian technique that quantitatively compares models with different
numbers of free parameters.  A description of this technique along
with model comparisons is given in Section \ref{modcomp}.  Section
\ref{system} discusses the impact of systematic distance
uncertainties.  Finally, we discuss two anomalous galaxies in
\S~\ref{gals}, and consider the validity of these statistical
goodness-of-fit measures in the presence of known systematic errors in
our underlying assumptions. Color gradients within individual galaxies
and unobserved gas point to small variations in their mass-light
ratios, at a level which naturally explains Milgrom fit discrepancies. 

\section{Data \& Modeling}\label{data}

BSK contains a detailed discussion of the data and fitting methods
used in this analysis, and the interested reader is pointed there for
a detailed description.  Briefly, we use a sample of 40 high surface
brightness galaxies with $I$-band images and Fabry-Perot
two-dimensional velocity maps.  These data have been provided by
Povilas Palunas and Ted Williams and a thorough description of the
acquistion and reduction of the data can be found in \citet{pw00}.  We
decompose the $I$-band images into disk and bulge components when
necessary.  The disk is assumed to be radially symmetric and
infinitesimally thin and the bulge is assumed to be spheroidal.

A mass model is determined once the mass-light ratios for the disk
(\mld) and the bulge (\mlb) are specified.  Despite evidence for
gradients in \ml\ factors \citep[\eg,][]{dej96}, we first consider
\ml\ values to be spatially constant; we relax this assumption in
\S~\ref{varyml}.  The $I$ subscript indicates mass-light ratios
determined from $I$-band photometry. We neglect any contribution from
gas, as we have no information about the radial distribution of gas in
the galaxies in our sample;  however, the gas in late-type galaxies
like those in our sample is likely to contribute $\lesssim 10\%$ of
the circular speed at any radius \citep{bro92,verh97}.  Neglecting it
will cause a slight systematic overestimate of \mld\ and can lead to
models mildly underpredicting circular speeds in the outer parts of
galaxies.  It could be argued that omitting the gas contribution will
cause our stars-only Milgrom models to describe the data worse than
other models simply because we are not including all relevant
information; \ie, our Milgrom models are born at a disadvantage.
However, our Milgrom models do not show any systematic trends
indicating that large amounts of gas would improve the models; the
Milgrom models do not systematically underpredict the rotation speeds
at the edge of the kinematic data.  We discuss the impact of this
omission in more detail for two interesting galaxies in \S~\ref{gals}.

The disk-bulge mass model determines the Newtonian circular speed
$v_c(\theta)$ as a function of angular separation from the center of
the galaxy. This is converted to a circular speed $v_c(r)$ as a
function of distance from the center of the galaxy, using the angular
diameter distance to the galaxy given by the Hubble distance
(systematic errors associated with this simple distance assumption are
discussed in Sec.~\ref{distances}).  Equation~\ref{mondeq} converts
this Newtonian circular velocity to a Milgrom-relation circular
velocity.  The Newtonian disk-bulge mass model is also combined with
the standard dark matter halo models (Eqs.~\ref{nfwrho} \ref{isorho},
and \ref{plrho}) to give the total rotation curves in these cases.

The rotation curve inferred from the model mass distribution is
compared with the actual galaxy rotation as measured by the
Fabry-Perot maps.  Throughout this paper, we fix the position angle
$\phi_0$, inclination angle $i$, and systemic velocity $v_s$ of each
galaxy at the values deduced from earlier fits of axisymmetric flow
patterns to our kinematic data \citep{bs03}.  We have found the values
of these parameters that minimize $\chi^2$ between the two-dimensional
velocity map and a planar axisymmetric flow pattern.  The fit
therefore yields a set of mean orbital speeds, and their
uncertainties, at equally-spaced radii which we show as the ``data''
in subsequent figures illustrating rotation curves.  Since we fit
axisymmetric mass models, we could, in principle, fit only this
one-dimentional rotation curve, which represents the speed averaged in
annular bins, provided the weights are computed correctly.  However,
the full two-dimensional maps have several important advantages,
including clear identification of outlying pixels and identification
of non-axisymmetric velocity patterns (usually bars or spirals) which
cause systematic errors in rotation curves obtained from
one-dimensional velocity measurements.  Of course, identifying
non-axisymmetric patterns is only half the battle.  We are looking
towards collecting data for a sample chosen with minimal spiral and/or
bar structures that will involve fewer systematic effects and will
allow better model discrimination.

\section{Milgrom Fits}\label{mondfits}

The mass model incorporating disk and bulge components gives a
Newtonian circular velocity $v_c(r)$ and acceleration $a_N(r)$.
Equation~\ref{mondeq} then gives a total acceleration $a(r)$; equating
these accelerations with centripetal accelerations give us the Milgrom
circular speeds. Note that for the galaxies considered here, the
simple acceleration approximation in Eq.~\ref{mondeq} gives
essentially the same answer as the physically more realistic
Poisson-like Eq.~\ref{poisson} which is numerically more challenging
to solve.  Using previously determined projection angles and systemic
velocities \citep{bs03}, which are held fixed, we project a
one-dimensional speed profile into a two-dimensional velocity map.
This model velocity map is compared to the data velocity map, pixel by
pixel.

A small number of pixels exhibit what are clearly anomalous
inferred velocities; these are almost all isolated pixels in outer
regions of galaxies, where the velocity is much different than the
smooth velocity field defined by the surrounding pixels. These points
are due to imperfect filtering of cosmic rays and unmodelled line
emission in the data analysis pipeline, which spuriously fits the
pixels as having high velocity and low dispersion.  To reduce the
influence of these non-normal errors, we find the best-fit values for
\mld\ and \mlb\ by minimizing the Tukey biweight parameter
\citep{press92},
\begin{equation}\label{defbiwt}
\biwt = \frac{1}{N}\sum_{i=1}^N
\cases{
z_i^2-\frac{z_i^4}{c^2}+\frac{z_i^6}{3c^4} & $|z_i| < c$,\cr
\frac{c^2}{3} & otherwise,}
\end{equation}
where $z_i\equiv (D_i-M_i)/\sigma_i$, $D_i$ and $M_i$ are the
respective measured and mass-model velocities in pixel $i$, $\sigma_i$
is the measurement standard error on $D_i$, $c$ is a constant, and $N$
is the number of data points.  Note that in the limit $c\rightarrow
\infty$, this reduces to the usual expression for reduced chi-squared
$\chisq=1/N \sum_{i=1}^N (D_i-M_i)^2/\sigma_i^2$.  While data values
that differ from the model prediction by $\la c\sigma_i/2$ contribute
to \biwt\ almost exactly as for the conventional \chisq, the
contribution from data values that are farther from the model
prediction by $c\sigma_i$ does not change as the parameters are
adjusted, and outlying data therefore do not influence the resulting
parameter values.  \citet{press92} recommend $c=6$, but non-circular
flow patterns in our galaxies suggest that a larger value would be
more appropriate; after some experimentation we find $c=10$ eliminates
the influence of the extreme outlying velocities while allowing the
great majority of the pixels to contribute to the fit with almost full
weight. Only a handful of pixels per galaxy image are ignored using
this procedure.

\subsection{Parameter Values and Rotation Curves}\label{parvals}

The best-fit \biwt, \mld\ and \mlb\ parameters, statistical
uncertainties ($\sigma$), and systematic uncertainties ($s$) for the
Milgrom mass models are given in Table~\ref{mltab}.   We also include
the ratio of the distance to the edge of the kinematic data $R_{\rm
k,max}$ to the disk scalelength $r_d$ for each galaxy.  On average,
the kinematic data extends to $\approx 4.5 r_d$ and covers the visible
disk.  The uncertainty calculations are discussed in the next
paragraph.  The average \mld\ is 1.9 and the average \mlb\ is 2.9,
with wide variation in both values.  Generally, the Milgrom model
values are lower than the Newtonian stars-only values; the ratio of
\ml\ between the Milgrom models described here and the Newtonian
stars-only models of BSK are 0.6 and 0.5 for the disk and bulge
components, respectively.

Statistical uncertainties have been calculated using Markov Chain
Monte Carlo (MCMC) techniques \citep[\eg,][]{chris01,kmj02} to map out
allowed regions of parameter space.  Projections of the boundary which
encloses the most likely 68\% of the models then provide the
statistical uncertainties.  The average statistical uncertainties are
0.08 for \mld\ and 0.3 for \mlb.  The quoted systematic uncertainties
in the fitted \mli\ values are due to the uncertainties in the
galaxy's inclination and position angle.  A previous paper
\citep{bs03} estimated systematic uncertainties in inclination and
position angles caused by spirals and other nonaxisymmetric structure
in the galaxies.  The best-fit value plus the estimated high and low
limits for each of the two projection angles yield nine different
combinations of inclination and position angle which give nine
separate estimates for \mli\ values.  We take the systematic
uncertainty in a \mli\ value to be half the range of these nine \mli\
values.  As in BSK, the systematic uncertainties dwarf the statistical
ones with an average $s_{\mld}$ of 0.5 and an average $s_{\mlb}$ of
0.8.

\begin{figure}
\plotone{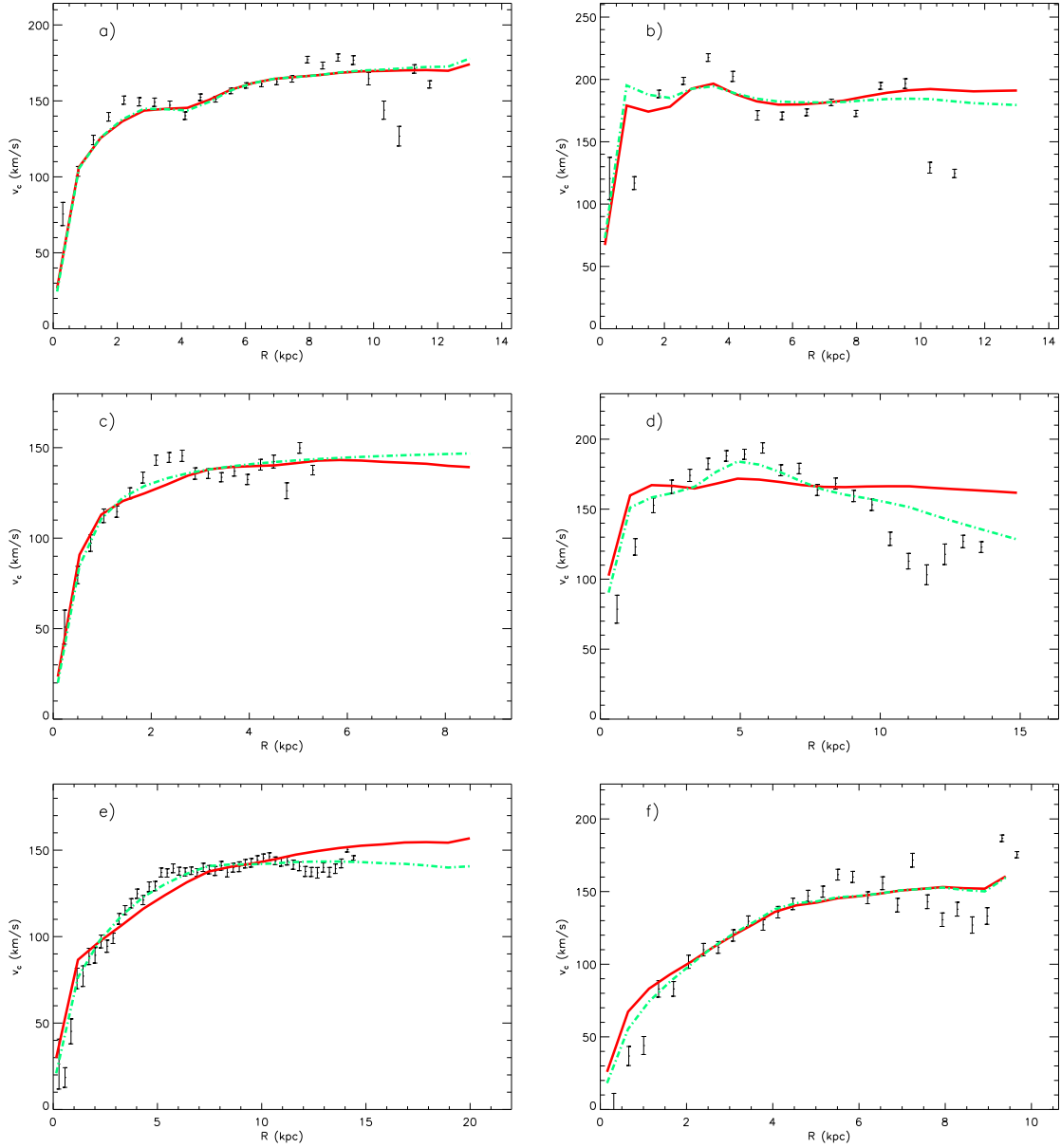}
\figcaption{Comparisons between Milgrom relation and pseudo-isothermal
halo rotation curves.  The solid lines are the Milgrom rotation curves
and the dot-dashed lines are the pseudo-isothermal halo curves.  The
error bars mark the circular speeds derived from the velocity map
during kinematic fitting \citep{bs03}.  Panels a) (ESO 322g76) and b)
(ESO 445g15) illustrate cases in which the Milgrom fit is preferred
over or comparable to the other models.  Panels c) (ESO 268g44) and d)
(ESO 438g08) show two cases where the Milgrom models are not favored
over any of the other models.  ESO 268g44 has substantial spiral
structure and ESO 438g08 appears to have an interacting companion.
Panels e) (ESO 323g42) and f) (ESO 382g06) are the axisymmetric
galaxies for which the Milgrom fit is never favored.  
\label{rcs}}
\end{figure}

Figure~\ref{rcs} illustrates the rotation curves of six galaxies
(assuming fixed Hubble distances).  The best-fit Milgrom models (solid
lines) are shown along with the data points derived from velocity
maps.  For comparison, we also show the PI halo models (dot-dashed
lines) from BSK.  The top two panels are for galaxies in which the
Milgrom fit is favored over the various BSK halo models.  The middle
set of panels show fits to galaxies with nonaxisymmetric structure in
which the Milgrom fit is less favored than the other BSK models.  The
bottom panels show curves for axisymmetric galaxies in which the
Milgrom fit is less favored than the BSK models.  These galaxies will
be discussed in more detail in \S~\ref{gals}.  One of the most
striking things about this figure is the close similarity bewteen the
Milgrom and halo curves; the differences in circular speeds between
the models shown are rarely much larger than 10 km/s.  Large
differences in likelihoods between the models are due to small
variations in model velocities over very large numbers of pixels in
the two-dimensional velocity maps.

\subsection{Color-\ml\ relations}\label{colors}

Stellar population synthesis models predict correlations between
galaxy color and \mld\ value, with red galaxies having larger \mld\
values than blue galaxies \citep[\eg,][]{bdj01,bell03}.  As in BSK, we
have created galaxy colors obtained from the ESO-LV aperture
magnitudes, as recorded in the NASA/IPAC Extragalactic Database (NED).
In principle, we should compare the \mld\ with the color of the disk
alone, but NED gives magnitudes of the disk and bulge combined.  The
bulge light in our mostly late-type galaxies is a small fraction,
typically $\leq 10\%$, of the total in the $I$-band, so any correction
for the bulge to the total color is likely to be small.  We have
corrected the NED magnitudes for galactic extinction using values from
\citet{schleg98} and for internal extinction using the prescriptions
of \citet{tully98}.  We estimate errors in our colors to be around
$0.15\;$mag.

In BSK, we found that the model \ml\ values for the galaxies in this
sample did not have any trend, and instead follow a flat distribution,
albeit with large scatter.  We present here the color-\mld\ relation
for the Milgrom models for these same galaxies.  Figure~\ref{lincomp}
shows the \mld\ value of each galaxy plotted versus its dereddened
color for Newtonian stars-only models (top frame) and Milgrom models
(bottom frame).  The error bars reflect the statistical and systematic
uncertainties added in quadrature.  Galaxies marked by squares have no
bulge components.  The solid lines in each frame represent the
correlation given in \citet{bell03}.  The dashed lines are the
best-fit linear fits to the data.  Contrary to the findings of BSK, we
see a trend in the Milgrom color-\ml\ plane similar to that expected
by stellar population models.  The slopes of these lines show some
discrepancy but there is still a large scatter in our points.

This sample of 40 galaxies contains 8 which belong to the Hydra
cluster.  The color-\mld\ plot of these galaxies is shown in
Figure~\ref{cluster}.  As before, the \citet{bell03} correlation is
shown as a solid line, while the best linear fit to the data (marked
by diamonds) is the dashed line.  This linear fit appears to
accurately describe the correlation present in the data with little
scatter.  The large offset between the two lines is surprising since
\citet{mcg04} found that MOND $\Upsilon_D$ values matched the
\citet{bell03} line remarkably well.  There are a number of possible
explanations of the discrepancy between the Hydra galaxies and the
\citet{bell03} correlation.  One possibility is that the offset is due
to errors in the adopted distances to these galaxies (see
\S~\ref{system} for a detailed discussion of the impact of distance on
MOND \ml\ values).  However, \citet{hud97} give a Fundamental
Plane-based, Malmquist bias corrected distance to the Hydra cluster of
$52.6\pm 2.5$ Mpc and the galaxies in this plot have adopted distances
that average to $52.6\pm 4.2$ Mpc.  Metallicities and initial mass
functions of galaxies have also been shown to strongly impact the
slope and intercept of the color-\mld\ correlation \citep{bdj01};
these possibilities would require further observations to explore in
more detail.

\section{Model Comparisons}\label{modcomp}

Previous studies have rarely attempted direct comparisons between
Milgrom fits and halo model fits. Here we compare the Bayesian
probabilities of halo models to Milgrom models, taking into account
the differing number of free parameters in the models. A brief summary
of the necessary Bayesian probability theory is given in the Appendix.
The model comparisons here are based on statistical fits only.
Evaluation of some systematic sources of error and their potential
impact on straightforward statistical model comparison are considered
in the following sections.

We are considering several different models with different numbers of
parameters and we would like to know which provides the most
economical description of the data.  Specifically, we have compared
our Milgrom fits to various halo fits (PI, NFW, and power law halos),
along with Newtonian stars-only fits for illustration (for a full
discussion of the various halo models, see BSK).  We have used the
following parameter ranges for (uniform) prior probability values:
(PI) $\Delta \rho_0=2$ M$_{\sun}/pc^3$, $\Delta r_c=20$ kpc; (NFW)
$\Delta c=20$, $\Delta v_{200}=300$ km/s; (power law) $\Delta v_0=300$
km/s, $\Delta \alpha=2$.  Altering the priors will change a given
model's probability, but other sensible choices do not change our
basic results.  Table \ref{probtab} lists the logarithms of the
probability ratios $\lambda$ of the BSK models to Milgrom fits.
$\lambda$ values $<0$ indicate that the Milgrom model is preferred.
The large ratios between some halo and Milgrom models are due to the
large numbers of data points in the two-dimensional velocity data,
resulting in very small statistical error bars and large likelihood
ratios between models with similar rotation curves.  A simpler model
comparison test based on Bayesian information criteria (BIC)
\citep{lid04} agrees with the results we present here. 
 
Surprisingly, the Newtonian stars-only model is strongly favored over
Milgrom for nine galaxies, and the models are comparable for two
others (ESO 317g41 and 445g39). In fact, the Newtonian stars-only
model is preferred over all other models for ESO 438g08 and 382g06.
Of these 11 galaxies, 5 have substantial bars, 3 have noticeable
spirals, and 1 has a close, possibly interacting, companion (ESO
438g08).  Since it is possible that these circumstances introduce a
bias into the model comparison, we do not focus our attention on these
galaxies.  The two remaining galaxies (ESO 323g42 and 382g06) appear
axisymmetric (no large bars or strong spiral structure).  These
galaxies are discussed in detail in \S~\ref{gals}.

For three galaxies, Milgrom models are preferred or are comparable to
all other models: ESO 322g76, ESO 445g15, and ESO 501g68.  ESO 445g15
and ESO 501g68 have bars while ESO 322g76 has moderately strong spiral
structure.  All 3 have velocity maps that extend at least 2.5 disk
scalelengths and their rotation curves have flattened by the last data
point. Of the 40 galaxies analyzed, these are the only galaxies for
which Milgrom is preferred over the PI halos. Milgrom models are
preferred over NFW halos in 7 cases. These results suggest that, in
general, the simplicity (fewer free parameters) of Milgrom models does
not statistically outweigh the ability of halo models to provide
better fits.

Due to its poor performance relative to the halo models, it is
tempting to conclude immediately that the Milgrom relation describes
galaxy rotation curves only by accident, undermining the strongest
reason for considering MOND as a modification of gravity.  However,
this conclusion would be premature without detailed consideration of
systematic uncertainties.  It is dfficult to find cases where the
poorer fit of the Milgrom model cannot be accounted for by reasonable
known systematic errors.

Milgrom models are highly constrained, and we have made several
implicit assumptions which have substantial effects on the rotation
curves: 1) all stellar motions are circular; 2) the visible stellar
light accounts for all of the mass in the galactic disk (neglecting,
in particular, any gas, which is known to be non-negligible in some
galaxies); 3) the mass-light ratio is constant throughout the galaxy;
4) the adopted inclination angle is correct; and 5) the galaxy
distance is given by the Hubble distance.  Not one of these
assumptions are exactly true, and in many galaxies these assumptions
are known to be incorrect at levels which would change the rotation
curve by an amount significantly larger than the statistical errors.
The sample considered here, in particular, was chosen to probe
large-scale velocities in the local universe, not because of its
particular suitability for galaxy dynamics; we have analyzed it
because of its large sample of high-quality two-dimensional velocity
measurements.  Milgrom models have essentially no ability to
compensate for any failure of these assumptions as they are determined
completely by the distribution of visible matter. Halo models, by
contrast, have two degrees of freedom which are {\it unconnected} to
the distribution of visible matter and otherwise unconstrained by
direct observation, and can therefore compensate to some degree for
any incorrect assumptions about the visible matter. In other words,
beyond the fact that halo models will generically provide better fits
because they have more parameters, they are also more flexible in
compensating for systematic errors, which is not reflected in purely
statistical comparisons. 

In the remainder of this paper, we consider in more detail some of the
important systematic errors which might affect the analysis presented
here. We then focus on two sample cases for which Newtonian fits are
statistically superior to either Milgrom or dark halo fits, and
discuss which possible explanations for the discrepancy are
reasonable. 

\section{Systematic Effects}\label{system}

\subsection{Distances}\label{distances}

A substantial source of uncertainty is the adopted distance to a
galaxy.  \ml\ values are inversely proportional to distance in both
Newtonian and Milgrom mass models.  Unfortunately, a significant
fraction ($\approx 50\%$) of the galaxies in this sample were
originally observed to study peculiar velocities in the region toward
the ``Great Attractor'' and may therefore have larger-than-usual
peculiar velocities and distance uncertainties.  \citet{bothun92} used
the Tully--Fisher relation to estimate distances to this subsample of
galaxies, finding differences between redshift distance and
Tully--Fisher distance of typically about 25\%.  While for Newtonian
(stars-only and halo) fits this simply introduces a scale uncertainty
in \ml\ values, Milgrom fits are subject to a further distance
dependence, since they depend on the dimensional parameter $a_0$.
Changing the distance to a galaxy changes the acceleration scale at a
given angular separation from the galaxy center, and thus changes the
inferred radius at which the non-Newtonian behavior sets in.  For
example, suppose the distance to a galaxy has been adopted at half its
actual distance.  Given angular separations now correspond to physical
distances that are half the actual values.  Since the accelerations
are inversely related to physical distances, the accelerations
increase.  These increases then push the location where acceleration
becomes comparable to $a_0$ farther out in the disk.  Moving a galaxy
closer forces the Milgrom fit to be closer to Newtonian, while
adopting a distance larger than the actual value increases the
discrepancy between Milgrom  and Newtonian fits.

While Milgrom fits are fairly sensitive to the adopted distance scale,
this fact can be exploited as a method for determining distances.  As
related in \citet{sm02}, changing the adopted distance to a galaxy
until a satisfactory Milgrom fit is obtained gives an independent
distance measurement that is often in accordance with Tully--Fisher
distances (which is not surprising, as the Tully--Fisher relation is
implied by the Milgrom relation).  We have investigated this
possibility by including the distance to galaxies as another free
parameter in Milgrom fits. For our data set, most fits of this kind
are unstable, leading to models with galaxies unreasonably nearby or
far away.  Part of the reason for this behavior is due to bars and
spirals.  Nonaxisymmetric features tend to lower rotational
velocities, making axisymmetric models overshoot the data, especially
in central regions.  In such a case, the fitting routine can better
match the data by moving the galaxy farther away, thereby lowering the
\ml\ values, making the baryonic component insignificant, and allowing
the non-Newtonian MOND behavior to dominate.  This is similar to the
behavior of some galaxies that are best fit with a halo alone, as
noted in BSK.  One of our galaxies (ESO 438g08) has a declining
rotation curve that is best fit by the Newtonian stars-only model and
returns a terrible fixed-distance MOND fit.  Once the distance is
allowed to change, the galaxy is moved unbelieveably closer so that
the fit can become more Newtonian.  If this galaxy's kinematics can be
shown to be due only to the matter within the galaxy, it would pose
quite a problem for MOND, as well as for any commonly considered dark
matter halo.  However, the galaxy has a nearby, possibly interacting,
companion; while we cannot definitively say that tidal interaction is
causing the declining rotation curve, neither can we rule it out.

In an attempt to remove the effects of aggravating central
asymmetries, we have excised pixels from the centers of the galaxy
images and re-fit the remaining data with a Milgrom model that has the
distance as an additional free parameter.  If the free parameter
distance varies by more than a factor of 3 from the Hubble distance we
used in the fixed distance fitting, we consider the distance fit
unstable.  We fit each galaxy 10 times, each time increasing the
number of excised pixels.  The first attempt utilizes all the pixels
and each successive trial removes an additional 5\% of the total
radial data range from the central regions.  Fifteen galaxies cannot
find acceptable distance fits for any amount of excised pixels.
Fourteen galaxies have acceptable fits using all the data.  Twenty-two
galaxies have at least three consecutive fits with acceptable
distances; however, the distances that these fits return never agree
with each other to better than 30\% of the Hubble distance.  So while
the adopted Hubble distance necessarily introduces a systematic
uncertainty that predominantly affects the Milgrom fits, we have not
been able to devise a stable method of determining more accurate
values directly from the dynamics of the galaxies. Rotation curve data
extending further into the flat portion of the rotation curve force a
stable distance determination from Milgrom fits; this is clearly an
important limitation of the data set considered here.

\subsection{Projection Angles}

Systematic uncertainties in projection angles substantially affect
derived \ml\ values (see the discussion in \S~\ref{parvals}).
However, the estimates of systematic uncertainties in the projection
angles appear to be reasonably behaved. Our sample contains 14
galaxies for which the uncertainty in \ml\ from the estimated
systematic uncertainty in projection angles is less than twice the
statistical uncertainty (see Table 1); in most of these cases the
systematic error is less than the statistical error. A cursory
examination of the model fits for these 14 galaxies shows that the
various halo models or the Milgrom model are preferred as the
best-fitting model in roughly the same fractions as for the full set
of galaxies.  More importantly, the galaxies of this subsample tend to
have $|\lambda| \la 10$, while the full sample contains several cases
with $|\lambda| \ga 100$. These large values are likely a signal that
the straightforward statistical comparison may be suspect, and that
systematic errors arising from projection angles or other sources are
contributing to the strong rejection of the Milgrom model.

\subsection{Nonaxisymmetry}

Many of the galaxies in the sample have bar and/or spiral features or
lopsided shapes. In these cases, the assumption of axisymmetry is
manifestly violated, although if the irregularities are weak (as is
generally the case) the effect on the rotation curve will be small.
Departures from regular elliptical (projected) isophotal shapes can
indicate recent merging or tidal interactions which substantially
distort the velocity field.  These distortions are evident in maps of
the differences between circular orbit models and kinematic data
\citep[\eg, see Figure 5 in][]{bs03}.  The effect of departures from
axisymmetry requires painstaking modelling \citep[\eg,][]{wsw01}, but
incisive comparisons of halo models with Milgrom fits should be
limited to galaxies with no gross departures from axisymmetry.  Our
sample contains roughly equal numbers of galaxies with strong (18) and
weak (22) spiral structure.  Seven galaxies have some evidence for a
bar; 5 are strong bars in galaxies with strong spiral structure, and 2
are short/weak bars in galaxies with weak spiral structure.  The
specific galaxies discussed in the following section have only weak
spiral structure.

\section{ESO 323g42 \& ESO 382g06}\label{gals}

MOND is vastly more falsifiable than dark matter hypotheses.  Since
the fitting parameters are few and understood more easily than the
usual halo parameters, any one galaxy that can not be fit by the
Milgrom relation could, in principle, refute the idea.  Most galaxies
in our sample have better halo fits than Milgrom fits; however,
as discussed in \S~\ref{distances}, these relative likelihoods
may be the result of systematic errors. 

A better way to rule out MOND is simply to find galaxies with poor
Milgrom fits in an absolute sense. This criterion also is hard to
define as measures of fit quality are also affected by systematic
errors.  One simple way to judge the quality of a Milgrom fit is in a
relative sense, by comparing with the Newtonian stars-only model.  If
the Milgrom relation provides an accurate description of galaxies,
then Newtonian fits should {\it never} be better than Milgrom fits
(they will be comparable if all of the measured velocities are close
enough to the galaxy center that they are still in the Newtonian
gravitational limit of the Milgrom force law). Many of the systematic
effects that the halo fits can compensate for will affect both Milgrom
fits and Newtonian fits equally, since both have little model freedom
to compensate for systematic errors. It is difficult to see, for
example, how an unaccounted-for distribution of gas could improve the
fit of a Newtonian model relative to a Milgrom model; it should affect
both equally and in the same direction relative to the measured
velocities. 

Finding galaxies whose Newtonian fits are clearly superior to Milgrom
fits can kill MOND as a fundamental theory. (Note that galaxies whose
Newtonian fits are superior to dark halo fits also pose severe
problems for dark matter models.) Here we focus on two galaxies, ESO
323g42 and ESO 382g06, whose Milgrom fits are worse than their
Newtonian stars-only fits. Their rotation curves and several model
curves are shown in Fig.~\ref{rcs2}(a) and (b), respectively.
 
In practice, the path to ruling out MOND is not so clear as systematic
uncertainties complicate the situation.  For these two galaxies, the
residuals from our kinematic fitting do not show organized patterns
associated with strong bars and/or spirals, and we believe these
galaxies are quite axisymmetric.  Any spiral structure present likely
perturbs circular speeds by $<10$ km/s, and the resulting small
rotation curve features would be similar in both Newtonian and Milgrom
fits.  Without detailed information about the gas distribution, it is
certainly possible that these Milgrom models are performing poorly
simply because they have incorrect input.  Further, the observed
discrepancies between the Milgrom models and the data have magnitudes
($\lesssim 10\%$) that could possibly be lifted by including the gas
\citep{bro92,verh97}.  We note that this is a question that can be
addressed with further observations.  

\subsection{ESO 323g42}

Figure~\ref{rcs2}(a) shows that the Milgrom relation fit (solid line)
is clearly a poorer approximation to the data than either the
Newtonian stars-only fit (dotted line) or the PI fit (dot-dashed line
and the best halo fit) over the entire radial range of the data,
$\approx 3.4$ disk scalelengths $R_d$ (the NFW curve has been omitted
for clarity but is very similar to the PI curve).  The systematic
projection angle uncertainties for this disk are quite small ($\approx
2^{\circ}$) and likely do not impact the fit strongly.  Unfortunately,
this galaxy is also one of the cases where no stable fit could be
found when the galaxy distance is left as a free parameter.  The
Hubble distance may not be the correct value, but there is no reliable
evidence to change it.  The discrepancy between the Milgrom curve and
the data is most evident where the model underpredicts $v_c$ near 5
kpc and overpredicts $v_c$ beyond 12 kpc.  A reasonable gas
distribution could rectify these discrepancies.  However, one problem
with adding significant amounts of gas to the model is that the model
\mld\ value necessarily has to decrease.  The \mld\ value in
Table~\ref{mltab} is 0.78, already somewhat lower than the average
value.  Looking at the PI model curve, it is apparent that departures
from axisymmetry are quite small, $< 5$ km/s.  Any mild spiral
structure that may be present is not greatly influencing the model.

\subsection{ESO 382g06}

Unlike the case of ESO 323g42, the Milgrom fit for ESO 382g06
basically agrees with the both PI and Newtonian stars-only fits
everywhere (out to the last point at $\approx 4.3 R_d$), except in the
very central region of the galaxy.  In that region, no model explains
the data well.  One possible explanation is that, despite lacking a
clear signature in either the photometric or kinematic residuals, the
center of ESO 382g06 harbors some kind of non-axisymmetric structure.
The rotation curve of ESO 382g06 decreases slightly towards the edge
of the optical disk, which is not so uncommon for late type spirals
\citep{bb01}.  Such behavior in the low-acceleration regime causes the
Milgrom relation fit to underperform the halo models, since it
unavoidably gives asymptotically flat rotation curves.  Note that this
discrepancy is directly tied to the shape of the rotation curve, and
will persist even including an undetected gas component.  The
projection angle uncertainties for this galaxy are about average for
this sample \citep[as determined in][]{bs03}, $\approx 5^{\circ}$.
However, it is unlikely that changing the projection angles will
change the slope of the outer part of the rotation curve.  Following
the procedure described in \S~\ref{system}, differing amounts of
central pixels have been removed from the fit, and in three trials
stable distance values have been found.  Unfortunately, these values
do not agree with each other very well and the fits give unphysically
small amounts of mass to the disk.  Consequently, we again find that
changing the adopted distance to the galaxy will not improve the
Milgrom fit.

\subsection{Variable \mld\ Fits}\label{varyml}

In Section~\ref{system}, we have discussed some of the observational
systematic uncertainties that could adversely influence the Milgrom
fits.  Here we consider the impact of a systematic modeling error due
to the fundamental assumption of a constant mass-light ratio \ml.
Milgrom fits will be particularly susceptible to this type of error
because of the deterministic dependence of the rotation curve on the
baryonic mass.  Halo models of any kind generally have sufficient
freedom to contort themselves so that mild \ml\ gradients leave the
quality of a halo fit unchanged.

We have performed Milgrom fits for three galaxies (ESO 322g76, ESO
323g42, and ESO 382g06) allowing the \mld\ values to vary radially;
\mlb\ values continue to be treated as spatially constant.
Specifically, we assume a linear relation,
\begin{equation}
\mld (r)=\mldz+m r,
\end{equation}
where $\mldz$ is the central disk \ml\ value and $m$ is the slope of
the relation in $(M_{\sun}/L_{\sun})$/kpc.  The rotation curves for
the variable \mld\ mass models are compared to the data points, the
constant \ml\ curves, and the PI halo curves in Figure~\ref{varycomp}.
The solid lines correspond to the variable \mld\ fits, the dot-dashed
lines are the PI halo fits, and the dotted lines are the constant \ml\
fits.

The fit for ESO 322g76 (Figure~\ref{varycomp}a) serves as a test case
since the constant \ml\ Milgrom fit is comparable or preferred to all
the other models.  We find $\mldz=1.30$, $\mlb=1.75$, and
$m=3.5\times10^{-3}$.  Since the slope is small and the \ml\ values
are nearly unchanged from their original values ($\mld=1.33$,
$\mlb=1.73$), it is not surprising that the fits are equally good (for
both fits, $\biwt=2.05$).  For this galaxy, with an acceptable Milgrom
model with constant \ml, the additional freedom does little to change
the previous fit.

This is not the case for the two galaxies that appear to have
problematic Milgrom fits.  Allowing for \mld\ gradients, ESO 323g42
(Figure~\ref{varycomp}b) is best fit with the parameters
$\mldz=\mlb=1.16$, $m=-4.0\times10^{-2}$.  The resulting \biwt\ value
is 1.54, roughly 90\% of the constant \mld\ value, making it
comparable to the constant \ml\ halo and Newtonian fits.  The change
in \mld\ value from the center to the edge of the kinematic data is
0.29, much larger than either uncertainty given in Table~\ref{mltab}.
This suggests that including the \ml\ gradient is truly improving the
fit for this galaxy.  The best-fit parameters for ESO 382g06
(Figure~\ref{varycomp}c) are $\mldz=\mlb=3.81$ and
$m=-2.5\times10^{-1}$.  This case has a slight decrease in \biwt\
value, from 2.48 with constant \mld\ values to 2.46.  In this case,
the \mld\ value changes from center to edge by 1.2, smaller than just
the statistical uncertainty alone in Table~\ref{mltab}.  While the
quality of the fit has definitely improved, it is not as clear in this
case that allowing a \ml\ gradient has significantly changed the
model.

Qualitatively, the \mld\ slopes for these two galaxies act in the
direction expected from observations of color gradients, with the
outermost regions having lower \mld\ values than the centers
\citep{dej96}.  Multi-band photometry would allow determination of the
actual color gradient, and whether this gradient is consistent with
the best fit value. 

ESO 323g42 and ESO 382g06 exemplify the difficulties associated with
falsifying the Milgrom relation.  While there are certainly hints of
problems with the zeroth-order models, plausible systematic effects
(distance, gas mass, \ml\ gradients) must be exhausted before a
definitive judgement can be handed down.  For Milgrom fits with
non-constant mass-light ratios, further observations can provide
consistency checks, unlike halo models where all of our ignorance is
folded into halo parameters and independent direct measurements of
individual halo shapes are generally not possible.

\section{Summary \& Conclusions}

We have investigated MOND mass models of 40 high surface surface
brightness spiral galaxies.  Using two-dimensional Fabry-Perot
velocity maps and $I$-band photometric images, best-fit bulge and disk
\ml\ values have been determined using the Milgrom relation
(Equation~\ref{mondeq}).  As in our previous work using dark matter
halo models \citep[][BSK]{bsk04}, we find that the systematic
uncertainties due to projection angle uncertainties are typically much
larger than the statistical uncertainties.  It should be kept in mind
that large systematic uncertainties could mask differences between
various models and make any statistical comparison questionable.

\begin{figure}
\epsscale{0.8}
\plotone{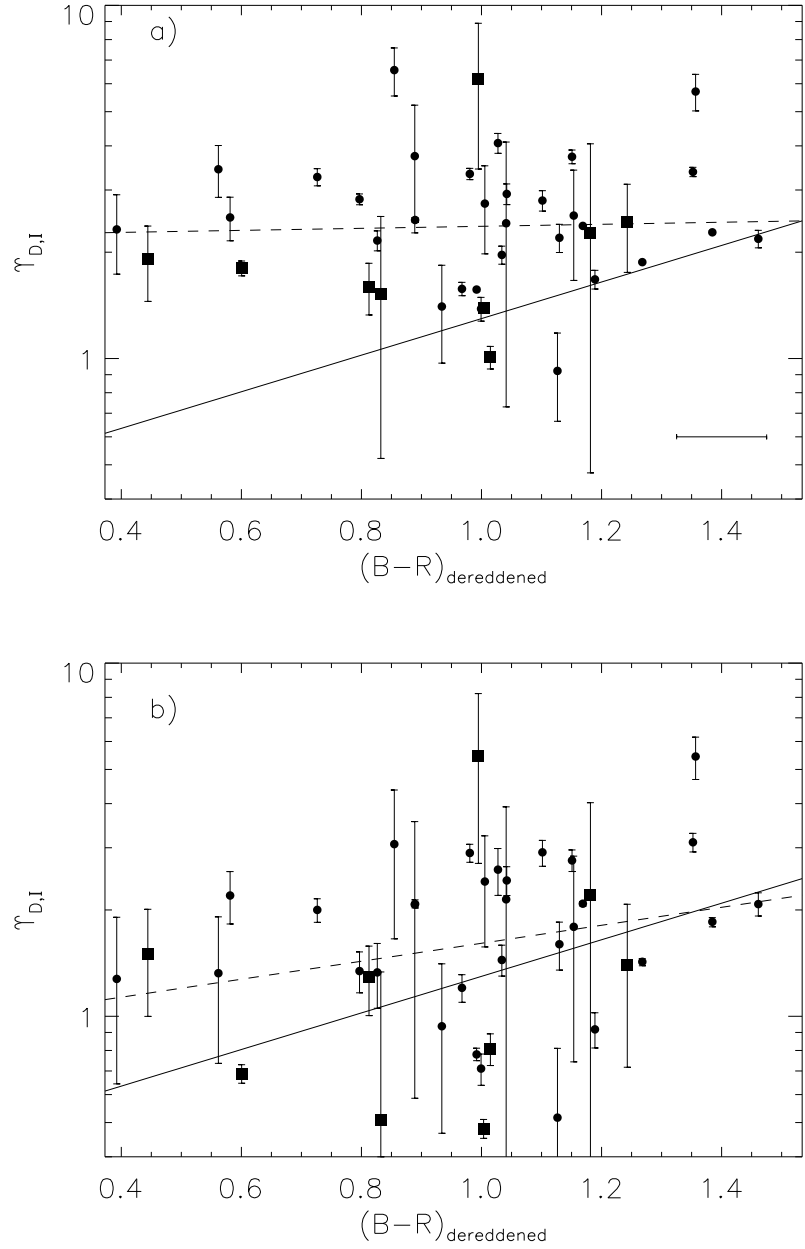}
\figcaption{The color-\mld\ correlations for the galaxies
in this sample assuming Newtonian stars-only models (a) and Milgrom
models (b).  The dashed line is the best linear fit to the points
and the solid line is the fit from \citet{bell03}.  The error bar is
the estimated uncertainty in galaxy color.
\label{lincomp}}
\end{figure}

\begin{figure}
\plotone{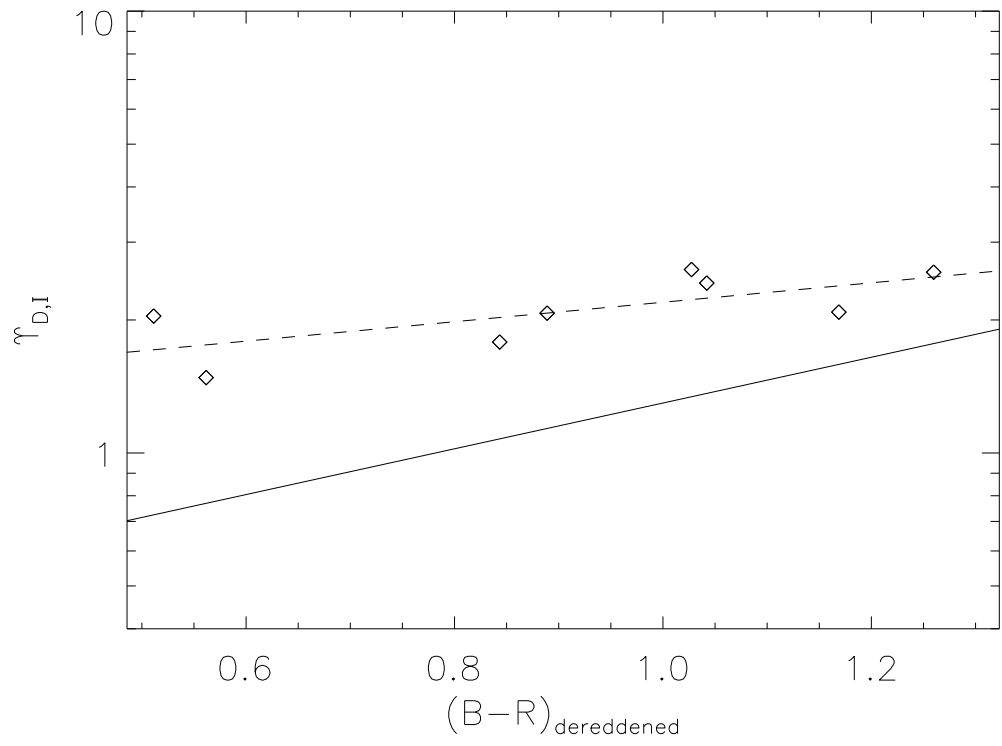}
\figcaption{The color-\mld\ correlations for the Hydra
cluster galaxies.  Again, the dashed line is the best linear fit to
the points and the solid line is the \citet{bell03} fit.
\label{cluster}}
\end{figure}

\begin{figure}
\plotone{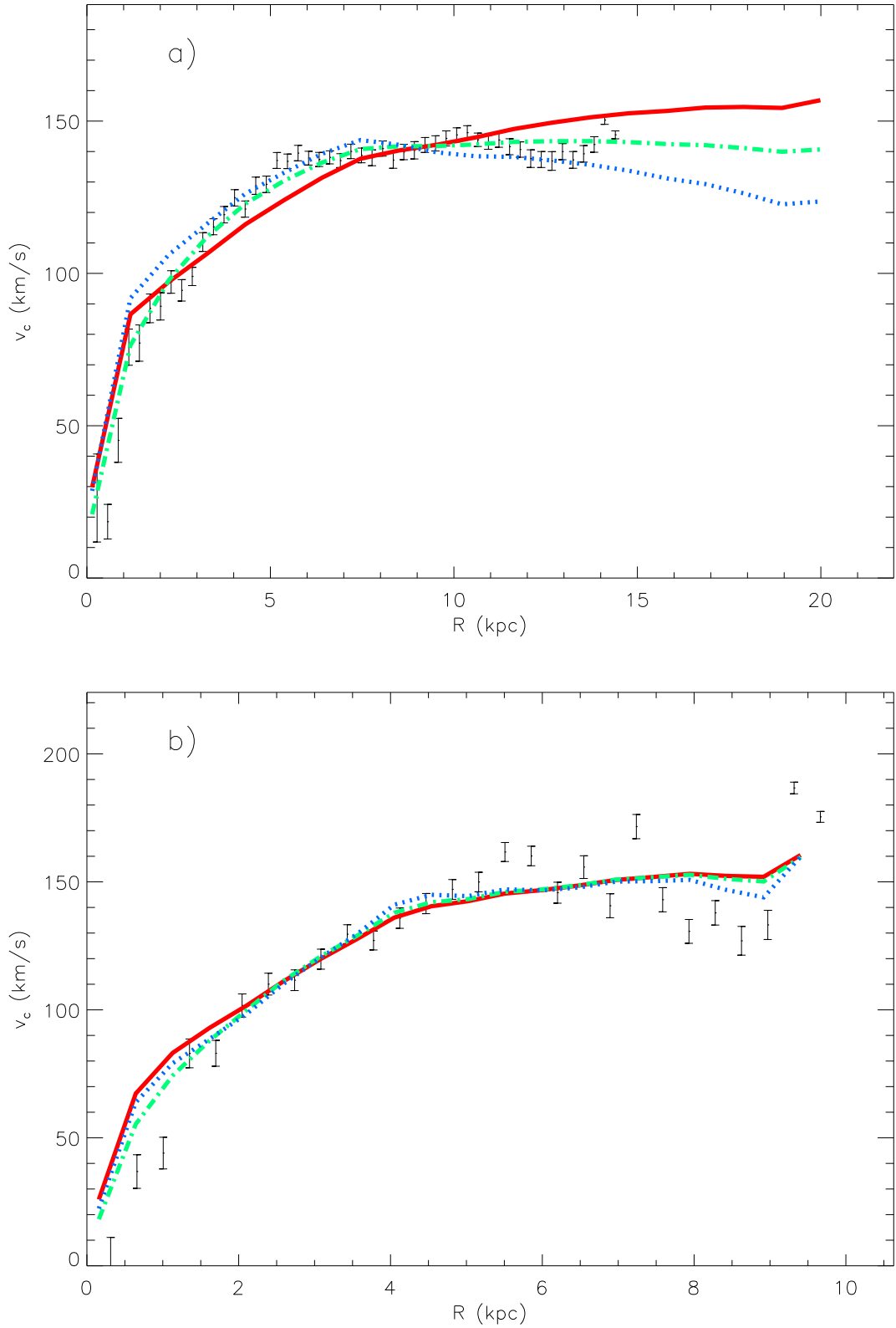}
\figcaption{Comparisons between Milgrom (solid), pseudo-isothermal
(dot-dashed), and Newtonian stars-only (dotted) rotation curves for
ESO 323g42 (a) and ESO 382g06 (b).  The points with error bars show
the circular speeds derived from the velocity map during kinematic
fitting \citep{bs03}.
\label{rcs2}}
\end{figure}

\begin{figure}
\scalebox{0.75}{
\plotone{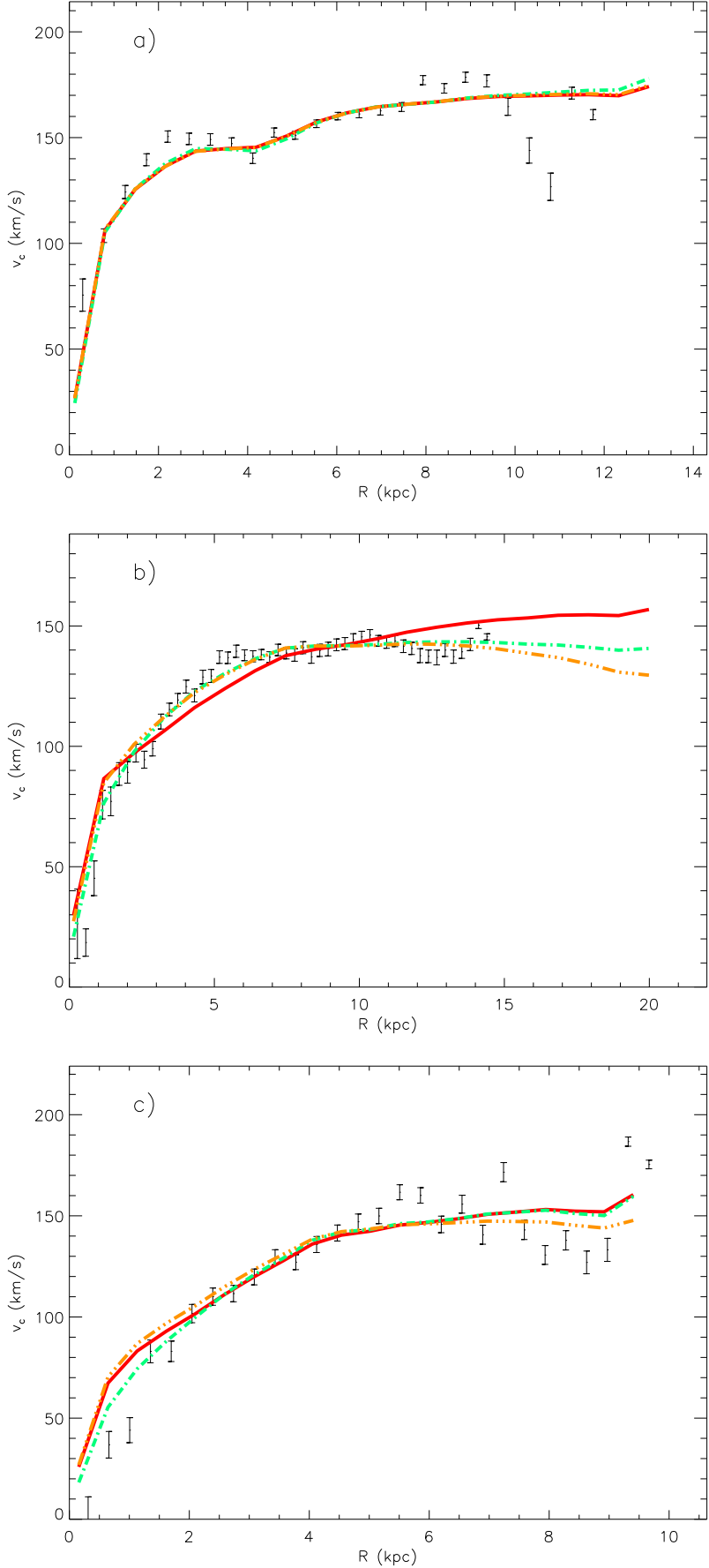}}
\figcaption{Comparisons between spatially-constant \mld\ Milgrom
(solid lines), pseudo-isothermal (dot-dashed lines), and variable
\mld\ Milgrom (triple-dot-dashed lines) rotation curves for ESO 322g76
(a), 323g42 (b), and ESO 382g06 (c).  Again, the points with error
bars denote the circular speeds derived from kinematic fitting of the
velocity map.
\label{varycomp}}
\end{figure}

We find that Milgrom model \mld\ values correlate with $B-R$ galaxy
color (Figure~\ref{lincomp}b), in line with population synthesis
models \citep{bdj01,bell03}.  While the correlation does not exactly
match the synthesis model trend and contains much scatter, it sharply
contrasts with the results of BSK where no trend was found
(Figure~\ref{lincomp}a).  For a subsample of 8 galaxies that lie in
the Hydra cluster, we find that they define a fairly tight \mld--color
correlation (Figure~\ref{cluster}).  Again, the correlation does not
agree with the trends from population synthesis models in either slope
or intercept.  This contrasts with the findings in \citet{mcg04} which
show excellent agreement between Milgrom model and population synthesis
\ml--color relations.  We note that the galaxies used in \citet{mcg04}
are the same galaxies used to calibrate the population synthesis
prediction in \citet{bdj01}.  We feel confident that the disagreement
we see can not stem from distance errors, since the mean of the
distances to these galaxies is the same as an independent
determination using the Fundamental Plane \citep{hud97}.  It could be
argued that bulge contamination skews our results, but these are all
late-type spirals with very little bulge.  It is difficult to imagine
that these bulges could cause such a radical shift in the correlation.
Recognizing that the common view is that IMFs are universal, one
possible explanation made plain by the model calculations of
\citet{bdj01} is that there are different IMFs for these two sets of
galaxies.  With only 8 galaxies, the evidence is weak
to challenge the standard view.  

Utilizing a Bayesian model comparison technique (\S~\ref{modcomp} and
\S~\ref{app}), we have compared Milgrom models to Newtonian
stars-only, pseudo-isothermal halo, Navarro-Frenk-White halo, and
power law halo models.  The results are presented in
Table~\ref{probtab}.  Overall, halo models are statistically favored
over Milgrom models.  At the same time, Milgrom models are generally
preferred to Newtonian stars-only models.  Unfortunately, this kind of
analysis takes statistical uncertainties only into account.  The
quality of the model fits is also influenced by systematic
observational uncertainties (distance, inclination, etc.; see
\S~\ref{system}).  The Milgrom models are also subject to systematic
modeling effects (like assuming constant mass-to-light ratios) that
are more easily absorbed in halo models.

We have also discussed two specific galaxies which appear to present a
problem for the Milgrom relation (\S~\ref{gals}).  Our investigation
suggests that nonaxisymmetry, projection angle uncertainties, or
distance errors are not responsible for the Newtonian stars-only fits
being better than the Milgrom fits.  These galaxies provide excellent
tests of the Milgrom relation since in this picture no galaxy should
show Newtonian behavior in the low-acceleration (large radius) regime.

With an eye towards explaining away the seemingly Newtonian behavior
in these two galaxies, we have performed additional fits allowing the
\ml\ of the disk to vary linearly with radius.  We find that for the
galaxy ESO 322g76, which is statistically well explained by the
Milgrom relation, the best-fit parameters of a constant \mld\ are
virtually unchanged by allowing \mld\ to vary.  However, for the two
problem galaxies ESO 323g42 and ESO 382g06 we find that 1) the
best-fit parameters of the varying \mld\ model are quite different
than the constant \mld\ model; 2) the quality of the fits improve to
the point of be comparable to halo fits; and 3) the predicted \mld\
variations agree qualitatively with observed variations.  It is
possible that a more quantitative match could be recovered with the
proper observations. 

In conclusion, while we find little statistical evidence to support
the claim that Milgrom models explain the behavior of this sample of
galaxies more economically than do dark matter models, there are
several sources of systematic uncertainty that complicate the
situation.  Given these systematic uncertainties (distance, gas
content, \ml\ gradients), it is difficult to see how Milgrom mass
models are of lower quality than dark matter halo models.  However, if
one believes that stellar population models accurately describe
galaxies, then correct models of the stellar components of real
galaxies should reveal trends expected from such models.  This is one
piece of evidence in favor of Milgrom over dark matter models; Milgrom
model \ml\ values show more of a trend with color than do dark matter
model \ml\ values for the same galaxies.

Dynamical analyses which purport to rule out the Milgrom relation must
be done on carefully selected galaxies for which systematic errors in
dynamical assumptions are minimized. It is easy to demonstrate
superior dynamical fits by dark matter halos than by the Milgrom
relation, but on closer examination it is clear that in the majority
of cases, the discrepancy can be explained by reasonable assumptions
about particular known systematic errors. 

\acknowledgements 
The authors gratefully acknowledge the assistance and advice of
Povilas Palunas, Ted Williams, Stacy McGaugh, and an anonymous
referee.  Ted Williams provided the imaging and Fabry-Perot data on
which this paper is based. We also thank Tad Pryor, Eric Peng, Pat
C\^ot\'e, and Laura Ferrarese for several helpful discussions.  This
work was supported by NASA grant NAG5-10110, and AK has been partly
supported by NSF grant AST-0546035  while finishing this work.  We
have used the NASA/IPAC Extragalactic Database (NED), which is
operated by the Jet Propulsion Laboratory, California Institute of
Technology, under contract with the National Aeronautics and Space
Administration. 

\appendix

\section{Bayesian Statistics Primer}\label{app}

The application of Bayesian reasoning to statistical problems in
astronomy has become much more widespread in recent years, and
numerous textbook treatments exist. Here we give only a brief summary
of Bayesian model comparison used in this paper, and point the
interested reader to \citet{jaynes03,brett90}.

Bayesian statistics is based on Bayes' theorem, 
\begin{equation}\label{bayes}
P(M|DI)=\frac{P(M|I)P(D|MI)}{P(D|I)}.
\end{equation}
$P(M|DI)$ is read as ``the probability that $M$ is true given $D$ and
$I$''.  Let us make this concrete.  Say that proposition $M$ is a
given model, proposition $D$ is the data fit by the model, and
proposition $I$ is any initial information we have about the system.
The left hand side of Equation~\ref{bayes} is then the probability
that a model is true given the data and prior information.  If we have
one data set, two models, and the same prior information, we can write
Equation~\ref{bayes} for each model $M_j$ and $M_k$.  Since $D$ and
$I$ are the same in both equations, we can divide the two to cancel
$P(D|I)$.  We now have a ratio of the probabilities of the models
being correct,
\begin{equation}\label{probrat}
\frac{P(M_j|DI)}{P(M_k|DI)}=\frac{P(M_j|I)P(D|M_jI)}
{P(M_k|I)P(D|M_kI)}.
\end{equation}
Following \citet{jaynes03} we have that,
\begin{eqnarray}
P(D|MI) & = &\int d\mathbf{\Theta} P(D|\mathbf{\Theta} MI)
P(\mathbf{\Theta}|MI) \nonumber \\
 & = & \int d\mathbf{\Theta} L(\mathbf{\Theta}) P(\mathbf{\Theta}|MI),
\end{eqnarray}
where $\mathbf{\Theta}$ are the $(1...N)$ free parameters, $L$ is the
likelihood of a model, and $P(\mathbf{\Theta}|MI)$ is the prior
probability.  If we normalize the likelihood by its maximum
value $L_{\rm max}$, we have that $P(D|MI)=L_{\rm max}W$, where $W$ is
the Ockham factor,
\begin{equation}
W\equiv \int d\mathbf{\Theta} \frac{L(\mathbf{\Theta})}{L_{\rm max}}
P(\mathbf{\Theta}|MI).
\end{equation}

In order to calculate the Ockham factor we need to decide what to use
for the prior parameter probability.  In the absence of any specific
knowledge of the parameters, we choose a top-hat prior, \ie, a constant
value $p$ in a specified range.  Since the prior probability must be
normalized, we can easily calculate what $p$ is.  $\int
P(\mathbf{\Theta}|MI) d\mathbf{\Theta}=1$ implies that
$p=1/(\Delta\Theta_1 \Delta\Theta_2...\Delta\Theta_N)$, where
$\Delta\Theta_j$ is the range of a given free parameter.  With this 
prior we can write,
\begin{eqnarray}
P(D|MI) & = & L_{\rm max} p \int d\mathbf{\Theta}
\frac{L(\mathbf{\Theta})}{L_{\rm max}} \nonumber \\
 & = & L_{\rm max} p V(\Omega),
\end{eqnarray}
where $V(\Omega)$ is just the integral of the normalized likelihood
over all parameter space and is related to the volume of allowed
parameter space.  We are now ready to rewrite the right-hand 
side of Equation~\ref{probrat},
\begin{equation}\label{mcomp}
\frac{P(M_j|DI)}{P(M_k|DI)}=\frac{P(M_j|I)}{P(M_k|I)} \frac{L_{{\rm
max},j}}{L_{{\rm max},k}} \frac{p_j}{p_k}
\frac{V_j(\Omega_j)}{V_k(\Omega_k)}.
\end{equation}
Assuming our prior information leaves us equally ignorant about both
models, the first term on the right-hand side of Equation~\ref{mcomp}
is 1.  We are left with an expression that is simple and sensible.  If
$M_j$ provides a better fit (\ie, has a smaller \biwt) than $M_k$,
then the likelihood ratio (the second term) is large, thus pushing the
ratio in favor of $M_j$, as one would expect.  If the two models have
the same free parameters, then the third ratio relating the
parameter ranges is 1 and the last ratio is $\approx 1$.
We are left with the obvious answer that the model with the highest
likelihood is the more probable model.  However, the real power of
this approach emerges when you compare models with different numbers of 
parameters.  For concreteness, let us say that $M_j$ has two free
parameters $\alpha$ and $\beta$ and model $M_k$ has two additional
parameters $\gamma$ and $\delta$.  We know $L_{\rm max}$ for each
model, and we assume that $L_{{\rm max},k}<L_{{\rm max},j}$.  We also
know that, $p_j=1/(\Delta\alpha \Delta\beta)$ and $p_k=1/(\Delta\alpha
\Delta\beta \Delta\gamma \Delta\delta)$, so the $p$ ratio term is
$\Delta\gamma \Delta\delta$.  Since $M_k$ has the larger number of
parameters, let us suppose that $V_j(\Omega_j)>V_k(\Omega_k)$.  The
likelihood ratio is $<1$, the $V$ ratio is $>1$, and the $p$ ratio is
dependent on the ranges that we believe two of the parameters fall in.
The Bayesian model probability ratio does not necessarily choose the model
with the highest likelihood: it determines if the benefit of
adding more fitting parameters balances the cost of making the model
more complicated.

\begin{deluxetable}{lcccccc} 
\tablewidth{0pt} 
\tablecaption{Best-fit \biwt values, \ml\ values (with statistical
uncertainties $\sigma$), and systematic uncertainties ($s$) for
Milgrom models.  The last column contains the ratio between the
kinematic data's radial extent $R_{\rm k,max}$ and the disk
scalelength $r_d$.\label{mltab}} 
\tablehead{\colhead{Galaxy} & \colhead{$\biwt$} & 
\colhead{\mld $^{+\sigma}_{-\sigma}$
\tablenotemark{a}} & 
\colhead{$s_{\mld}$} & 
\colhead{\mlb $^{+\sigma}_{-\sigma}$} & 
\colhead{$s_{\mlb}$} & \colhead{$R_{\rm k,max}/r_d$}}
\startdata 
ESO 215g39 & 1.48 & 
1.28 $^{0.07}_{0.09}$ & 0.63 & 
2.79 $^{0.32}_{0.25}$ & 0.57 & 3.35\\ 
ESO 216g20 & 4.16 & 
2.08 $^{0.07}_{0.07}$ & 0.14 & 
1.31 $^{0.30}_{0.28}$ & 0.37 & 5.20\\ 
ESO 263g14 & 1.93 & 
0.81 $^{0.01}_{0.02}$ & 0.08 & 
\nodata $^{ }_{ }$ & \nodata & 4.64\\ 
ESO 267g29 & 5.42 & 
2.14 $^{0.04}_{0.02}$ & 1.77 & 
2.14 $^{0.04}_{0.02}$ & 1.90 & 3.81\\ 
ESO 268g37 & 4.11 & 
1.40 $^{0.03}_{0.03}$ & 0.64 & 
\nodata $^{ }_{ }$ & \nodata & 3.78\\ 
ESO 268g44 & 2.05 & 
1.20 $^{0.07}_{0.06}$ & 0.08 & 
3.09 $^{0.27}_{0.34}$ & 0.15 & 4.45\\ 
ESO 317g41 & 4.58 & 
2.08 $^{0.04}_{0.04}$ & 0.04 & 
1.43 $^{0.12}_{0.10}$ & 0.04 & 5.09\\ 
ESO 322g36 & 4.53 & 
0.94 $^{0.02}_{0.03}$ & 0.47 & 
1.09 $^{0.09}_{0.08}$ & 0.38 & 3.71\\ 
ESO 322g42 & 3.67 & 
0.48 $^{0.02}_{0.02}$ & 0.02 & 
\nodata $^{ }_{ }$ & \nodata & 3.66\\ 
ESO 322g44 & 1.61 & 
0.71 $^{0.03}_{0.04}$ & 0.06 & 
0.72 $^{1.19}_{0.72}$ & 2.33 & 4.05\\ 
ESO 322g45 & 2.16 & 
1.43 $^{0.02}_{0.04}$ & 0.02 & 
1.43 $^{0.02}_{0.04}$ & 0.02 & 2.11\\ 
ESO 322g76 & 2.05 & 
1.33 $^{0.05}_{0.06}$ & 0.27 & 
1.73 $^{0.15}_{0.15}$ & 0.19 & 4.47\\ 
ESO 322g77 & 3.25 & 
3.11 $^{0.13}_{0.13}$ & 0.13 & 
2.08 $^{0.51}_{0.52}$ & 0.23 & 6.33\\ 
ESO 322g82 & 4.53 & 
1.34 $^{0.06}_{0.05}$ & 0.17 & 
7.99 $^{0.38}_{0.44}$ & 1.72 & 4.12\\ 
ESO 323g25 & 2.81 & 
2.41 $^{0.03}_{0.03}$ & 0.83 & 
15.56 $^{1.49}_{1.26}$ & 8.63 & 4.75\\ 
ESO 323g27 & 2.83 & 
1.44 $^{0.07}_{0.06}$ & 0.13 & 
5.10 $^{0.26}_{0.31}$ & 0.05 & 4.25\\ 
ESO 323g39 & 4.55 & 
0.51 $^{0.03}_{0.04}$ & 0.84 & 
\nodata $^{ }_{ }$ & \nodata & 3.09\\ 
ESO 323g42 & 1.66 & 
0.78 $^{0.03}_{0.03}$ & 0.01 & 
1.60 $^{0.28}_{0.21}$ & 0.08 & 4.50\\ 
ESO 323g73 & 3.10 & 
1.29 $^{0.02}_{0.02}$ & 0.29 & 
\nodata $^{ }_{ }$ & \nodata & 5.11\\ 
ESO 374g03 & 3.29 & 
0.69 $^{0.01}_{0.02}$ & 0.04 & 
\nodata $^{ }_{ }$ & \nodata & 3.84\\ 
ESO 375g02 & 2.08 & 
1.60 $^{0.05}_{0.06}$ & 0.24 & 
1.64 $^{0.32}_{0.36}$ & 0.43 & 5.35\\ 
ESO 381g05 & 2.75 & 
5.46 $^{0.07}_{0.09}$ & 2.86 & 
\nodata $^{ }_{ }$ & \nodata & 3.73\\ 
ESO 382g06 & 2.48 & 
3.07 $^{0.66}_{0.86}$ & 1.11 & 
3.42 $^{0.61}_{0.47}$ & 0.72 & 4.03\\ 
ESO 435g26 & 11.98 & 
1.79 $^{0.03}_{0.03}$ & 1.05 & 
3.20 $^{0.54}_{0.48}$ & 1.38 & 6.04\\ 
ESO 437g04 & 2.91 & 
2.42 $^{0.03}_{0.02}$ & 0.23 & 
2.42 $^{0.03}_{0.02}$ & 0.23 & 4.51\\ 
ESO 437g31 & 1.07 & 
2.07 $^{0.04}_{0.04}$ & 1.49 & 
2.07 $^{0.04}_{0.04}$ & 1.65 & 3.74\\ 
ESO 438g08 & 2.32 & 
0.52 $^{0.05}_{0.05}$ & 0.29 & 
3.66 $^{0.44}_{0.33}$ & 1.53 & 4.71\\ 
ESO 438g15 & 2.60 & 
0.92 $^{0.03}_{0.03}$ & 0.10 & 
1.66 $^{0.07}_{0.07}$ & 0.09 & 6.62\\ 
ESO 439g18 & 4.14 & 
2.20 $^{0.02}_{0.03}$ & 0.37 & 
2.20 $^{0.02}_{0.03}$ & 0.37 & 4.46\\ 
ESO 439g20 & 1.86 & 
2.90 $^{0.02}_{0.03}$ & 0.17 & 
2.90 $^{0.02}_{0.03}$ & 0.42 & 3.98\\ 
ESO 444g47 & 1.69 & 
2.00 $^{0.04}_{0.05}$ & 0.15 & 
2.00 $^{0.04}_{0.05}$ & 0.15 & 4.95\\ 
ESO 445g15 & 3.03 & 
2.91 $^{0.16}_{0.18}$ & 0.17 & 
1.58 $^{0.19}_{0.17}$ & 0.13 & 5.17\\ 
ESO 445g19 & 2.31 & 
1.85 $^{0.04}_{0.05}$ & 0.03 & 
1.12 $^{0.36}_{0.31}$ & 0.08 & 3.69\\ 
ESO 445g39 & 4.22 & 
5.44 $^{0.37}_{0.41}$ & 0.63 & 
2.89 $^{0.40}_{0.36}$ & 0.76 & 4.11\\ 
ESO 446g01 & 2.23 & 
1.50 $^{0.02}_{0.02}$ & 0.50 & 
\nodata $^{ }_{ }$ & \nodata & 3.97\\ 
ESO 501g01 & 2.33 & 
1.32 $^{0.03}_{0.04}$ & 0.44 & 
1.32 $^{0.03}_{0.04}$ & 0.44 & 2.76\\ 
ESO 501g68 & 2.13 & 
2.60 $^{0.22}_{0.25}$ & 0.31 & 
4.76 $^{0.21}_{0.21}$ & 0.15 & 3.84\\ 
ESO 502g02 & 1.74 & 
2.08 $^{0.02}_{0.02}$ & 0.02 & 
2.08 $^{0.02}_{0.02}$ & 0.02 & 4.50\\ 
ESO 509g80 & 3.75 & 
2.76 $^{0.07}_{0.05}$ & 0.19 & 
3.17 $^{0.18}_{0.17}$ & 0.14 & 4.83\\ 
ESO 569g17 & 1.67 & 
2.20 $^{0.03}_{0.03}$ & 1.83 & 
\nodata $^{ }_{ }$ & \nodata & 5.72\\ 
\enddata 
\tablenotetext{a}{\ml\ values have units 
(M$_{\sun}$/L$_{\sun}$).} 
\end{deluxetable} 

\begin{deluxetable}{lcccc} 
\tablewidth{0pt} 
\tablecaption{Logarithms of probability ratios for the various models.
The columns list the relative probabilities of the the various models
to Milgrom models.  Negative values indicate that the Milgrom model is
preferred.  The largest positive value for a galaxy comes from the
model (other than Milgrom) that best describes the data.
\label{probtab}} 
\tablehead{\colhead{Galaxy} & 
\colhead{$\lambda_{\rm stars-only}$} & 
\colhead{$\lambda_{\rm PI}$} & 
\colhead{$\lambda_{\rm NFW}$} & 
\colhead{$\lambda_{\rm power-law}$}}
\startdata 
ESO 215g39 & -44.66 & 40.26 & 
32.27 & 23.78 \\ 
ESO 216g20 & -32.43 & 9.06 & 
3.92 & 5.61 \\ 
ESO 263g14 & -49.46 & 7.16 & 
4.38 & 0.29 \\ 
ESO 267g29 & -47.16 & 27.47 & 
24.76 & 25.42 \\ 
ESO 268g37 & -170.16 & 33.57 & 
17.81 & 1.97 \\ 
ESO 268g44 & 3.21 & 8.95 & 
6.34 & 0.45 \\ 
ESO 317g41 & 0.05 & 4.24 & 
-15.56 & -17.21 \\ 
ESO 322g36 & -144.46 & 16.31 & 
15.70 & 14.20 \\ 
ESO 322g42 & -81.50 & 49.17 & 
11.56 & 12.90 \\ 
ESO 322g44 & -153.87 & 7.76 & 
3.93 & 5.06 \\ 
ESO 322g45 & -39.33 & 14.59 & 
11.65 & 13.18 \\ 
ESO 322g76 & -17.42 & 0.13 & 
-4.39 & -1.50 \\ 
ESO 322g77 & -3.96 & 8.18 & 
7.07 & 8.02 \\ 
ESO 322g82 & 158.84 & 162.29 & 
163.86 & 158.06 \\ 
ESO 323g25 & 59.15 & 177.19 & 
195.25 & 113.44 \\ 
ESO 323g27 & -4.76 & 21.32 & 
17.77 & 12.86 \\ 
ESO 323g39 & -55.35 & 20.97 & 
12.84 & 11.75 \\ 
ESO 323g42 & 36.22 & 79.43 & 
64.07 & 61.33 \\ 
ESO 323g73 & -94.97 & 53.64 & 
51.04 & 52.33 \\ 
ESO 374g03 & -78.04 & 71.89 & 
30.20 & 0.59 \\ 
ESO 375g02 & -84.92 & 21.97 & 
21.95 & 20.80 \\ 
ESO 381g05 & -128.24 & 215.20 & 
216.79 & 212.01 \\ 
ESO 382g06 & 10.70 & 9.89 & 
6.46 & 7.60 \\ 
ESO 435g26 & 126.10 & 182.98 & 
192.48 & 126.17 \\ 
ESO 437g04 & -153.84 & 207.50 & 
171.75 & 158.56 \\ 
ESO 437g31 & -67.01 & 13.52 & 
-3.45 & -2.09 \\ 
ESO 438g08 & 77.78 & 76.00 & 
73.35 & 74.73 \\ 
ESO 438g15 & -92.73 & 43.05 & 
39.63 & 41.19 \\ 
ESO 439g18 & -114.20 & 73.18 & 
63.88 & 35.15 \\ 
ESO 439g20 & -65.38 & 35.96 & 
24.74 & 3.19 \\ 
ESO 444g47 & -96.68 & 14.98 & 
2.95 & 5.09 \\ 
ESO 445g15 & -79.48 & -9.18 & 
-13.31 & -13.43 \\ 
ESO 445g19 & -67.36 & 7.28 & 
13.19 & 13.93 \\ 
ESO 445g39 & -0.26 & 7.52 & 
0.03 & 1.37 \\ 
ESO 446g01 & -219.48 & 4.36 & 
0.25 & 1.24 \\ 
ESO 501g01 & -66.08 & 31.50 & 
8.88 & -11.19 \\ 
ESO 501g68 & -4.56 & -1.74 & 
-4.70 & -2.35 \\ 
ESO 502g02 & -80.73 & 16.47 & 
12.31 & 5.60 \\ 
ESO 509g80 & 56.56 & 169.81 & 
166.45 & 90.89 \\ 
ESO 569g17 & 5.67 & 43.13 & 
46.70 & 9.39 \\ 
\enddata 
\end{deluxetable} 

\end{document}